\documentclass[prl,showpacs,twocolumn]{revtex4}%
\input{psfig.sty}
\usepackage{amsmath}
\usepackage{graphicx}
\usepackage{amsfonts}
\usepackage{amssymb}%

\begin{document}
\title{Quantum interference of electromagnetic fields from remote quantum memories}
\date{\today }
\author{T. Chaneli\`{e}re, D. N. Matsukevich, S. D. Jenkins, S.-Y. Lan, R. Zhao, T. A. B. Kennedy, and A. Kuzmich}
\affiliation{School of Physics, Georgia Institute of Technology,
Atlanta, Georgia 30332-0430} 
\begin{abstract}
We observe quantum, Hong-Ou-Mandel, interference of fields produced
by two remote atomic memories.  High-visibility interference is
obtained by utilizing the finite atomic memory time in four-photon
delayed coincidence measurements. Interference of fields from remote
atomic memories is a crucial element in protocols for scalable
generation of multi-node remote qubit entanglement.
\end{abstract}
\maketitle

Proposed approaches to scalable quantum information networks and
distributed quantum computing involve linear optical elements and
single-photon detectors \cite{knill,duan,chaneliere}. Photoelectric
detection events signal entanglement creation and, by postselection,
eliminate undesirable components of the electromagnetic field. While
postselection has a residual negative effect on the scaling of the
overall efficiency of quantum information protocols, this can be
offset by quantum memory, a resource which provides the capability
to perform quantum state transfer from matter to light and vice
versa, as demonstrated with cold atomic ensembles
\cite{matsukevich,matsukevich1,matsukevich2}. These also act as
sources of entangled photon pairs, with quantum memory enabling
user-controlled delays between the photons.

In order to distribute entanglement over a network configuration we
must connect entangled elements at remote sites. This may be
achieved by interfering photons, produced at these sites, on a
beamsplitter followed by coincident photoelectric detection. The
anticorrelation of coincidence counts is the signature of
Hong-Ou-Mandel interference (HOM), whereby single photons
simultaneously incident at two input ports of a beamsplitter both
exit in one or other of the output ports \cite{hom,mandel-wolf}. For
distinct, remote quantum memory elements, HOM is a possible method
for entanglement connection operations that scale efficiently with
the number of elements. Several remarkable demonstrations of HOM
using parametric down-conversion (PDC) have been reported (see Refs.
\cite{mandel-wolf,mandel,zeilinger} and references therein). It has
also been observed using photon pairs generated locally by a single
source - a quantum dot \cite{santori}, an atom \cite{legero}, and an
atomic ensemble \cite{thompson}. Recently HOM has been demonstrated
with two (a) neutral atoms \cite{beugnon} and (b) ions \cite{maunz},
in each case separated by a few microns.

\begin{figure}[htp]
\begin{center}
\leavevmode  \psfig{file=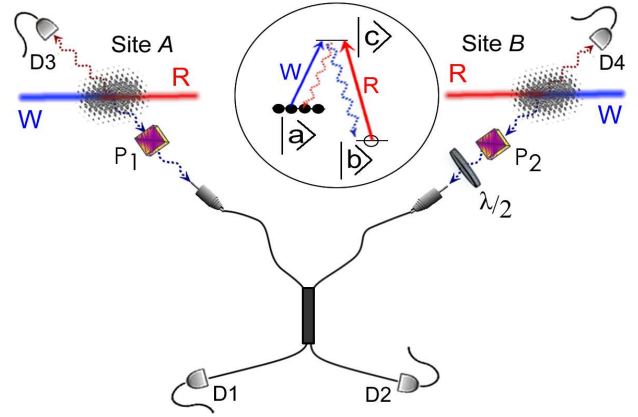,width=3.3in}
\end{center}
\caption{Schematic showing Raman scattering of write pulses (W) at
sites A and B with signal fields collected by polarizers $P_1$ and
$P_2$ and optical fiber beamsplitter and directed towards detectors
D1, D2. A half-wave plate ($\lambda /2$) may be inserted at site B
to rotate light polarization. Raman scattering of delayed read
pulses produces idler fields detected at D3, D4. The inset shows the
atomic level structure and the write- and read-induced Raman
processes.}\label{Fig1}
\end{figure}

In this Letter we report HOM from remote sources (cold atomic
ensembles) with quantum memory, located in adjacent laboratories and
separated by 5.5 m (Fig. 1). In a particular ensemble signal photons
are generated by Raman scattering of a write laser pulse with
temporal profile $\varphi\left( t\right) $ (normalized to unity
$\int dt~\left\vert \varphi\left( t\right) \right\vert ^{2}=1$),
whose length is much greater than the ensemble dimensions. For an
unpolarized ensemble of $N$ atoms interacting with an off-resonant
vertically ($\mathbf{e}_{V}=-\mathbf{\hat{z}}$) polarized write
field propagating in the $y$-direction, the interaction picture
Hamiltonian for the system includes a term representing Raman
scattering into the detected signal mode defined by an optical
fiber, and described by \cite{jenkins}
\begin{eqnarray}
\hat{H}\left(  t\right)  =i\hbar\chi\varphi\left(  t\right) \left(
\cos\eta~\hat{\psi}_{H}^{\dag}\left( t\right) \hat {s}_{H}^{\dag}+
\sin\eta~\hat{\psi}_{V}^{\dag}\left( t\right)
\hat{s}_{V}^{\dag}\right) + h.c.
\end{eqnarray}
The annihilation operator for the Raman scattered field emitted from
the ensemble is given by $\hat{\psi }_{\lambda}\left(  t\right)  $,
$\lambda =H,V$. These field operators obey the usual free-field,
narrow bandwidth bosonic commutation relations $\left[
\hat{\psi}_{\lambda}\left( t\right)
,\hat{\psi}_{\lambda^{\prime}}^{\dag}\left( t^{\prime}\right)
\right] =\delta_{\lambda,\lambda^{\prime}}\delta\left(
t-t^{\prime}\right) $. The emission of H- or V-polarized signal
photons creates correlated atomic spin wave excitations with
annihilation operators $\hat{s}_{H,V}$ given by, $\hat{s}_{V}
=-\hat{s}_{0}$, and
\begin{eqnarray}
\hat{s}_{H}  = \cos \theta  \hat{s}_{-1}- \sin\theta \hat{s}_{+1},
\end{eqnarray}
where $ \cos^2 \theta  = \sum_{m}p_{m}X_{m,-1}^{2}/
\sum_{\alpha=\pm1}\sum_{m}p_{m}X_{m,\alpha}^{2} $ and the spherical
vector components of the spin wave are given by
\[
\hat{s}_{\alpha}\equiv\sum_{m=-F_{a}}^{F_{a}}\frac{\sqrt{p_{m}}X_{m,\alpha}%
}{\sqrt{\sum_{m}p_{m}\left\vert X_{m,\alpha}\right\vert ^{2}}}\hat
{s}_{m,\alpha}.%
\]%
An atom is prepared in the state $\left\vert a,m\right\rangle $ with
probability $p_{m}=1/\left(  2F_{a}+1\right)  $ for an unpolarized
ensemble. Here $|a\rangle$ and $|b\rangle$ are the ground hyperfine
levels and $|c\rangle$ is the excited level associated with the
$D_1$-line involved in the Raman process, with total angular momenta
$F_a, F_b$ and $F_c$, respectively, and $X_{m,\alpha}\equiv
C_{m~0~m}^{F_{a}~1~F_{c}}C_{m-\alpha ~\alpha~m}^{F_{b}~1~F_{c}}$ is
a product of Clebsch-Gordan coefficients. The spin wave is defined
in terms of the $\mu$-th atom transition operators
$\tilde{\sigma}_{a,m;~b,m^{\prime}}^{\mu}( t)$ and the write
$\phi_{w}(\mathbf{r})$ and signal $\phi_{s}(\mathbf{r})$ mode
spatial profiles (normalized to unity in their respective transverse
planes)
\begin{equation}
\hat{s}_{m,\alpha}\equiv \frac{i\bar{A}}{\sqrt{p_{m}N}%
}\sum_{\mu=1}^{N}\tilde{\sigma}_{a,m;~b,m-\alpha}^{\mu}(t)
e^{i(\mathbf{k}_{s}-\mathbf{k}_{w})  \cdot
\mathbf{r}_{\mu}}\phi_{s}(\mathbf{r}_{\mu})
\phi_w^{\ast}(\mathbf{r}_{\mu}),
\end{equation}
where $\mathbf{k}_{s}$ and $\mathbf{k}_{w}$ are the detected signal
and write beam wavevectors, respectively, and $\bar{A}$ is the
effective overlap area of the write beam and the detected signal
mode \cite{jenkins,kuzmichkennedy}.

The dimensionless parametric coupling constant $\chi$ is given by
\begin{equation}
\chi \equiv \frac{2d_{cb}d_{ca}}{\Delta
}\frac{\sqrt{k_{s}k_{w}n_{w}N}}{\hbar\epsilon_{0}\bar{A}}
\sqrt{\sum_{m}p_{m}(X_{m,0}^{2}+\sum_{\alpha=\pm
1}X_{m,\alpha}^{2}/2)},
\end{equation}
where  $\Delta=ck_{w}-\left(  \omega_{c}%
-\omega_{a}\right)  $ is the write laser detuning from the $c
\leftrightarrow a$ transition, $d_{cb}$ and $d_{ca}$ are reduced
dipole matrix elements for the $c \leftrightarrow b$ and $c
\leftrightarrow a$ transitions, $n_{w}$ is the average number of
photons in the write pulse, and the parametric mixing angle $\eta$
is given by \cite{matsukevich1,jenkins}
\begin{equation}
\cos^{2}\eta=\frac{\frac{1}{2}\sum_{\alpha=\pm1}\sum_{m}p_{m}X_{m,\alpha}^{2}}{\sum_{m}p_{m}X_{m,0}^{2}+\frac{1}{2}\sum_{\alpha=\pm1}\sum_{m}p_{m}X_{m,\alpha}^{2}}.
\label{b}
\end{equation}
The interaction picture Hamiltonian also includes terms representing
Rayleigh scattering, as well as the Raman scattering into undetected
modes. One can show however that these commute with the signal
Hamiltonian to order $O\left( 1/\sqrt{N}\right)  $ and that they
commute with  $\hat{\psi}_{\lambda}$ and $\hat{s}_{\alpha}$. As a
result, the interaction picture density operator for the signal-spin
wave system (tracing over uncollected modes) is given by $
\hat{\rho}=\hat{U} \hat{\rho}_{0}\hat{U}^{\dag } $, where
$\hat{\rho}_{0}$ is the initial state of the unpolarized ensemble
and vacuum electromagnetic field and the unitary operator $\hat{U}$
is given by
\begin{eqnarray}
\hat{U} =\exp\left(  \chi\cos\eta
\hat{a}_{H}^{\dag}\hat{s}_{H}^{\dag}+
 \chi\sin\eta \hat{a}_{V}^{\dag}\hat{s}_{V}^{\dag}
- h.c. \right)
\end{eqnarray}
with the discrete signal mode bosonic operator
$\hat{a}_{\lambda}\equiv$$\int dt~\varphi^{\ast}\left(  t\right)
\hat{\psi }_{\lambda}\left(  t\right)$,
where $\lambda =H$ or $V$. In effect the multi-mode, multi-particle
system reduces to a discrete mode parametric interaction.

Since $\hat{a}_{\lambda}$ commutes with the Rayleigh scattering and
undetected Raman scattering Hamiltonians, the Heisenberg picture
solutions are given by
\begin{align*}
\hat{a}_{H}^{\left(  out\right)  } & =\hat{U}^{\dag}
\hat{a}_{H}^{\left(  in\right) }\hat{U} \\
& =\cosh\left(  \chi\cos\eta \right)  ~\hat{a}_{H}^{\left( in\right)
}+\sinh\left(  \chi\cos\eta \right) \hat{s}_{H}^{\left( in\right)
\dag},
\end{align*}%
\begin{align*}
\hat{a}_{V}^{\left(  out\right)  } & =\hat{U}^{\dag}\hat{a}_{V}^{\left(  in\right)  }\hat{U}\\
& =\cosh\left(\chi\sin\eta \right)  ~\hat{a}_{V}^{\left( in\right)
}+\sinh\left(\chi\sin\eta \right)  \hat{s}_{V}^{\left(  in\right)
\dag}.
\end{align*}
These solutions allow calculation of the photoelectric detection
signal for an atomic quantum memory element. In this work, two
magneto-optical traps (MOTs) of $^{85}$Rb, A and B, located in
adjacent laboratories, serve as the basis for distant quantum
memories (Fig.1). The corresponding Raman scattered signal field
operator for the detected mode with spatiotemporal mode $\varphi
_{A[B]} (t-z_{A[B]}/c)$ is given by
\[
\hat{E}_{\lambda,A}^{\left(  +\right)  }=\sqrt{\frac{\hbar k_{s}}%
{2\epsilon_{0}}}e^{-ick_{s}\left(  t-\frac{z_{A}}{c}\right)
}\phi_{s,A}\left(  \mathbf{r}\right)  \varphi _{A}\left(
t-\frac{z_{A}}{c}\right) \hat{a}^{\left(  out\right)
}_{\lambda,A},\] with a similar expression for ensemble B.

The fields from A and B are combined on a beamsplitter $R+T=1$,
where $R$ and $T$ are its reflectance and transmittance, and the
fields $\hat{E}_{\lambda,1}$, $\hat{E}_{\lambda,2}$ in the output
ports $1$ and $2$ are incident on detectors D1 and D2, respectively.
We employ vertically (V) polarized write beams, derived from the
same laser, and detect the horizontally (H) polarized signal fields,
which are passed through polarizing cubes prior to the beamsplitter.

 The corresponding cross-correlation function
 $G_{\|}^{(12)}(t,t+\tau) \equiv \langle \hat E^-_{H,1}(t)\hat E^-_{H,2}(t+\tau )\hat E^+_{H,2}(t+\tau
 )\hat E^+_{H,1}(t)\rangle$ exhibits the HOM effect:
\begin{eqnarray}
&&G_{\|}^{(12)}(t,t+\tau) =  \label{a}\\
&&{\mathcal E}_A^2 {\mathcal E}_B^2| T
\varphi_A(t+\tau)\varphi_B(t) - R\varphi_B(t+\tau)\varphi_A(t)|^2 s_{A}^2 s_{B}^2 \nonumber \\
&+&2RT \left( {\mathcal E}_A^4 |\varphi_A(t+\tau)\varphi_A(t)|^2
s_{A}^4 + {\mathcal E}_B^4 |\varphi_B(t+\tau)\varphi_B(t)|^2 s_{B}^4
\right) \nonumber
\end{eqnarray}
where $s_{A[B]}\equiv \sinh( \chi _{A[B]} \cos \eta)$, and $
\mathcal{E}_{A[B]}=\sqrt{\hbar ck_{s}/(2\epsilon_{0})}|\phi _{s,A[B]
}\left(  \mathbf{r}\right)| $. The first, HOM, term on the
right-hand side of Eq.(\ref{a}) exhibits two photon interference and
can be understood in terms of conventional single-photon
interference conditioned on the first photoelectric detection at
time $t$ \cite{legero}. For zero delay $\tau =0$ and a symmetric
beamsplitter $R=T=1/2$, this term gives zero contribution even for
$\varphi _A \neq \varphi _B $. Alternatively, for $\varphi _A =
\varphi _B $ it vanishes for arbitrary $\tau$. However,
$G_{\|}^{(12)}(t,t+\tau)$ does not vanish completely due to
contributions from multiphoton signal excitations (second term in
Eq. (\ref{a})). To quantify the degree of the HOM effect, the
following benchmark measurement is performed. We insert a half-wave
plate into the path of the signal field from ensemble B, rotating
its polarization from H to V, thus nullifying the HOM effect.
Quantitatively, in this case the corresponding correlation function
$G_{\perp}^{(12)}(t,t+\tau) $ is given by Eq.(\ref{a}), but now
without the interference contributions (proportional to the product
$RT$) in the HOM term.

In our experiment magneto-optical traps (MOTs) of $^{85}$Rb are used
to provide optically thick atomic ensembles at sites A and B
(Fig.1). The ground levels $\{|a\rangle;|b\rangle \}$ correspond to
the $5S_{1/2},F=\{3,2\}$ levels of $^{85}$Rb, and the excited level
$|c\rangle$ represents the $\{5P_{1/2},F=3\}$ level of the $D_1$
line at 795 nm. For a linearly polarized write field we observe that
the signal field is nearly orthogonally polarized, consistent with
the theoretical value of $\cos^2 \eta = 91/122$, Eq.(\ref{b}).

In order to generate indistinguishable signal wavepackets from the
two atomic memories, we produce their respective {\it write} fields
by splitting a single pulse and directing the outputs into identical
100 m long optical fibers. The two Raman-scattered signal fields
produced at A and B are passed through polarizing cubes to select
the H-components and coupled into the ends of a fiber-based beam
splitter. The outputs of the latter are connected to single-photon
counting modules D1 and D2. A half-wave plate is inserted into the
path of signal field B (Fig. 1) which allows us to vary the relative
(linear) polarization of the detected fields. This allows us to
detect parallel polarizations ($||$), which exhibit the HOM effect,
and orthogonal polarizations ($\perp$), which do not.

Particular care is taken to eliminate possible sources of spectral
broadening. Magnetic trapping fields are switched off after atomic
collection and cooling, and the residual ambient field is
compensated by sets of Helmholtz coils. All trapping and cooling
light fields are switched off during data acquisition. The trapping
light is shut off about 10 $\mu$s before the repumping light,
preparing unpolarized atoms in level $|a\rangle$.

In Fig. 2 we show the measured ratio of the photoelectric
coincidence rates ${\mathcal{R}_{\|}}/{\mathcal{R}_{\perp}}$, which
are integrated over the duration of the write pulses. Our
measurements exclude Rayleigh scattering on the {\it write}
transition by means of frequency filtering. The experimental ratio
$\mathcal{R}_{\|}/\mathcal{R}_{\perp}$ is compared to the ratio of
integrated correlation functions $\int\int dt d \tau
G_{\|}^{(12)}(t,t+\tau)$ and $\int\int dt d\tau
G_{\perp}^{(12)}(t,t+\tau)$, assuming identical wavepackets
$\varphi_A = \varphi_B$. We observe scatter in the data beyond the
deviations due to photoelectron counting statistics. These indicate
the level of systematic drifts encountered over several hours of
data acquisition.

The photoelectric coincidences arise from the signal field
excitation pairs produced (I) one excitation from each ensemble;
(II) both excitations from ensemble A; (III) both excitations from
ensemble B. The HOM visibility of $V \equiv 1-
\mathcal{R}_{\|}/\mathcal{R}_{\perp} = 1/3$ reflects the deleterious
effects of contributions (II) and (III). These are relatively large
because in the limit of weak excitation the spin wave-signal state
is dominated by the vacuum contribution. By detecting the presence
of a spin wave atomic excitation in each ensemble, these
contributions could be substantially suppressed, and the HOM
visibility $V \rightarrow 1$ in the limit that the excitation
probability $p_1 \rightarrow  0$.

\begin{figure}[t]
\begin{center}
\leavevmode  \psfig{file=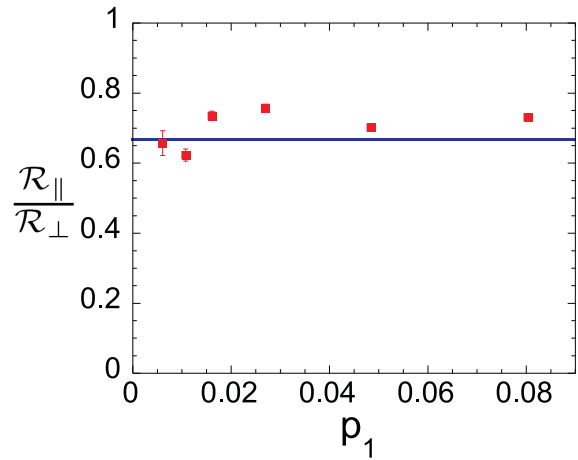,width=3.0in}
\end{center}
\caption{ Ratio of measured two-fold coincidence rates for the
$\perp$ and $\|$ configurations). The parameter $p_1\equiv
(N_1+N_2)/N_T$ (averaged over the $\perp$ and $\|$ cases). Here
$N_1(N_2)$ is the number of photoelectric detections in detector
$D1(D2)$, $N_T$ is the number of experimental trials. Theoretically
it can be expressed as $p_1=\epsilon_{A} s^2_A + \epsilon_{B}
s_B^2$, where $\epsilon_{A}(\epsilon_{B}) \approx 0.05-0.07$ is the
overall probability to detect a signal photon from site A (site B)
by either D1 or D2. Scatter beyond the estimated Poissonian level of
uncertainty is consistent with systematic drifts in experimental
conditions, in particular the single count rates from each ensemble.
The solid line is our theoretical prediction based on Eq.(\ref{a}),
for $R=T=1/2$ and $\epsilon_{A} s^2_A =\epsilon_{B} s_B^2$.
}\label{TQ}
\end{figure}

We obtain high-visibility HOM fringes by means of a four-photon
delayed coincidence detection procedure. This involves conversion of
the spin wave excitation to an idler field by means of an incident
read laser pulse which follows the write pulse by a programmable
time delay $\delta t$ in the off-axis geometry \cite{balic}; $\delta
t$ is limited by the quantum memory coherence time $\tau _c$
\cite{chaneliere1}. By careful minimization of ambient magnetic
fields, $\tau _c > 30$ $\mu$s have been reported \cite{dspg}. In
this work we choose $\delta t = 100$ ns in order to maximize the
repetition rate of the protocol. The four-fold detection of the two
idler and two signal fields involves HOM of the two signal fields
and delayed coincidence detection of the idler fields at detectors
D3 and D4, as shown in Fig. 1.

\begin{figure}[btp]
\begin{center}
\leavevmode  \psfig{file=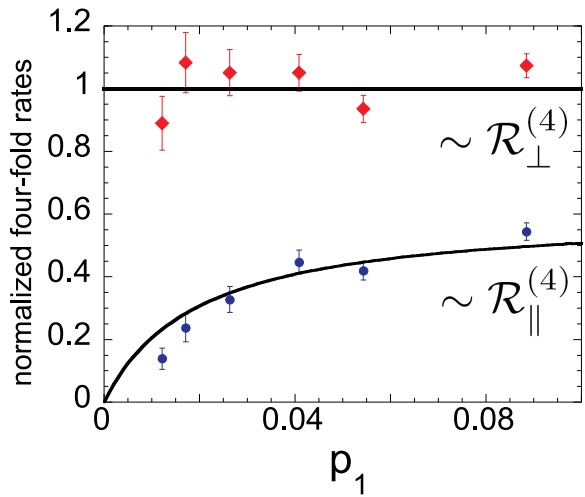,width=3.0in}
\end{center}
\caption{Integrated four-fold coincidence rates
$\mathcal{R}_{\|}^{(4)}/\mathcal{W}_{\perp}^{(4)}$ and
$\mathcal{R}_{\perp}^{(4)}/\mathcal{W}_{\perp}^{(4)}$ as a function
of $p_1$. Experiment, dots, theory, solid line, assuming identical
signal mode wavepackets from each ensemble. Uncertainties are based
on the statistics of the photon counting events.} \label{TQ}
\end{figure}

The four-fold coincidence rate is thus given by
\begin{eqnarray} \nonumber
\mathcal{R}_{\|}^{(4)} &\sim & s_A^2 s_B^2\left\{ (R-T)^2 (1+2
s_A^2)(1+2 s_B^2) \right. \\ &+& \left. 2RT \left(3s_A^4+3s_B^4
+2s_A^2 +2s_B^2 \right) \right\}.
\end{eqnarray}
We have again assumed identical wavepacket modes for both ensembles.

By inserting a half-wave plate into the path of the signal field
from ensemble B as before (rotating polarization from H to V), we
suppress the HOM interference contributions, such that the four-fold
coincidence rate becomes
\begin{eqnarray} \nonumber
\mathcal{R}_{\perp}^{(4)} &\sim& s_A^2 s_B^2\left\{ (R^2+T^2)(1+2
s_A^2)(1+2 s_B^2)\right. \\ &+& \left. 2RT \left(3s_A^4+3s_B^4
+2s_A^2 +2s_B^2 \right) \right\}.
\end{eqnarray}

In separate sets of measurements we recorded photoelectric events
with one, or other, of the two MOTs blocked, which allow us to
determine the expected level of four-fold coincidences for
orthogonal polarizations of the two signal fields
$\mathcal{W}_{\perp}^{(4)}$ (i.e., in the absence of HOM). In Fig. 3
we plot $\mathcal{R}_{\|}^{(4)}/\mathcal{W}_{\perp}^{(4)}$ and
$\mathcal{R}_{\perp}^{(4)}/\mathcal{W}_{\perp}^{(4)}$ along with the
corresponding theoretical predictions. HOM interference is
manifested in that $\mathcal{R}_{\|}^{(4)}/\mathcal{W}_{\perp}^{(4)}
\rightarrow 0$ as $p_1 \rightarrow 0$.  The highest observed
visibility $V\equiv 1-
\mathcal{R}_{\|}^{(4)}/\mathcal{W}_{\perp}^{(4)} \approx 0.86 \pm
0.03$. As the theory and the experimental data agree within the
statistical uncertainties, this indicates very good wave-packet
overlap of the signals produced by the remote ensembles.

In conclusion, we have demonstrated quantum interference of
electromagnetic fields emitted by remote quantum memory elements
separated by 5.5 m. Such high-visibility interference is an
important element towards scalable distributed entanglement, the
central resource in proposed quantum network and distributed quantum
computing systems \cite{knill,briegel,briegel1,lim,lim1,browne}.

We gratefully acknowledge illuminating discussions with M. S.
Chapman. This work was supported by NSF, ONR, NASA, Alfred P. Sloan
and Cullen-Peck Foundations.


\begin{thebibliography}{99}
\bibitem{knill} E. Knill, R. Laflamme, and G. J. Milburn, Nature (London) \textbf {409}, 46 (2001).
\bibitem{duan} L.-M. Duan {\it et al.}, Nature (London) \textbf{414}, 413 (2001).
\bibitem{chaneliere} T. Chaneli\`{e}re {\it et al.}, Phys. Rev. Lett. {\bf 96}, 093604 (2006).
\bibitem{matsukevich} D. N. Matsukevich and A. Kuzmich, Science {\bf 306}, 663 (2004); similar results were reported a year later
in C. W. Chou {\it et al.}, Nature (London), {\bf 438}, 828 (2005).
\bibitem{matsukevich1} D. N. Matsukevich {\it et al.}, Phys. Rev. Lett. {\bf 95}, 040405 (2005); similar results were reported a year later
in H. de Riedmatten {\it et al.}, Phys. Rev. Lett. {\bf 97}, 113603
(2006).
\bibitem{matsukevich2} D. N. Matsukevich {\it et al.}, Phys. Rev. Lett. {\bf 96}, 030405 (2006).
\bibitem{hom} C. K. Hong, Z. Y. Ou, and L. Mandel, Phys. Rev. Lett. {\bf 59}, 2044 (1987).
\bibitem{mandel-wolf}  L. Mandel and E. Wolf, \textit{Optical Coherence and Quantum Optics}, (Cambridge University Press, 1995).
\bibitem{mandel} L. Mandel, Rev. Mod. Phys. \textbf{71}, S274 (1999).
\bibitem{zeilinger} A. Zeilinger, Rev. Mod. Phys. \textbf{71}, S288 (1999).
\bibitem{santori} C. Santori {\it et al.}, Nature (London) \textbf{419}, 594 (2002).
\bibitem{legero} T. Legero, T. Wilk,  M. Hennrich, G. Rempe, and A. Kuhn, Phys. Rev. Lett.
{\bf 93}, 070503 (2004).
\bibitem{thompson} J. Thompson {\it et al.}, Science {\bf 313}, 74 (2006).
\bibitem{beugnon} J. Beugnon {\it et al.}, Nature (London) \textbf{440}, 779 (2006).
\bibitem{maunz}  P. Maunz {\it et al.}, quant-ph/0608047.
\bibitem{jenkins} S. D. Jenkins, Ph. D. Dissertation, Georgia Institute of
Technology (2006).
\bibitem{kuzmichkennedy} A. Kuzmich and T. A. B. Kennedy, Phys. Rev. Lett. {\bf 92}, 030407
(2004).
\bibitem{balic} V. Balic {\it et al.}, Phys. Rev. Lett. {\bf 94}, 183601 (2005).
\bibitem{chaneliere1} T. Chaneli\`{e}re {\it et al.}, Nature (London) {\bf 438}, 833
(2005).
\bibitem{dspg} D. N. Matsukevich {\it et al.},  Phys. Rev. Lett. {\bf 97}, 013601 (2006).
\bibitem{briegel}  H. J. Briegel, W. Duer, J. I. Cirac, and P. Zoller, Phys. Rev.
Lett. \textbf{81}, 5932 (1998).
\bibitem{briegel1}  H. J. Briegel and R. Raussendorf, Phys. Rev. Lett. \textbf{86},
910 (2001).
\bibitem{lim} Y. L. Lim, A. Beige, and L. C. Kwek, Phys. Rev. Lett. \textbf{95},
030505 (2005).
\bibitem{lim1} Y. L. Lim {\it et al.}, Phys. Rev. \textbf{73}, 012304 (2006).
\bibitem{browne}  D. E. Browne and T. Rudolph, Phys. Rev. Lett. {\bf 95}, 010501 (2005).
\end{thebibliography}
\end{document}